\documentclass[twoside,twocolumn,aps,prl,showpacs,groupedaddress]{revtex4}
\usepackage[latin9]{inputenc}
\usepackage{amsmath}
\usepackage{amssymb}
\usepackage{graphicx}
\usepackage{esint}

\makeatletter

\newcommand*\LyXThinSpace{\,\hspace{0pt}}

\@ifundefined{textcolor}{}
{%
 \definecolor{BLACK}{gray}{0}
 \definecolor{WHITE}{gray}{1}
 \definecolor{RED}{rgb}{1,0,0}
 \definecolor{GREEN}{rgb}{0,1,0}
 \definecolor{BLUE}{rgb}{0,0,1}
 \definecolor{CYAN}{cmyk}{1,0,0,0}
 \definecolor{MAGENTA}{cmyk}{0,1,0,0}
 \definecolor{YELLOW}{cmyk}{0,0,1,0}
}

\makeatother

\begin{document}

\title{Dynamical Quantum Phase Transitions in presence of a spin bath}

\author{Á. Gómez-León, and P.C.E. Stamp}

\affiliation{Department of Physics \& Astronomy, and Pacific Institute of Theoretical
Physics,\\
 University of British Columbia, Vancouver, Canada\\
 }

\date{\today}
\begin{abstract}
We derive an effective time independent Hamiltonian for the transverse
Ising model coupled to a spin bath, in the presence of a high frequency
AC magnetic field. The spin blocking mechanism that removes the quantum
phase transition can be suppressed by the AC field, allowing tunability
of the quantum critical point. We calculate the phase diagram, including
the nuclear spins, and apply the results to Quantum Ising systems
with long-range dipolar interactions; the example of $LiHoF_{4}$
is discussed in detail. 
\end{abstract}
\maketitle

\section{Introduction}

``Quantum phase transitions'' (QPT) take place between bulk equilibrium
phases in the zero-temperature ($T\rightarrow0$) limit. Hertz \cite{hertz}
showed that finite-$T$ thermodynamic and transport properties near
the zero-$T$ quantum critical point (QCP) should be determined solely
by the nature of the QCP itself. Classic examples are the Quantum
Ising system, and the paramagnetic/ferromagnetic (PM/FM) transition
in strongly-correlated conductors. However in real experimental systems
things are not so simple: in zero-$T$ PM/FM transitions, disorder
and 1st-order phase transitions often obscure the physics, and in
solid-state Quantum Ising systems ``environmental'' spin bath modes
\cite{spinB} can suppress the QCP entirely \cite{aeppli05,RonnowPRB}.
This is unfortunate, given the importance of Quantum Ising phenomenology
in so many areas of physics. There currently exists no good theory
of QPT in Ising systems in the presence of a spin bath; however, the
external control of the spin bath decoherence for qubits has been
studied from the perspective of Nuclear Magnetic Resonance, where
sequences of pulses are used to manipulate the coupling of qubits
to the environmental modes\cite{DynamDeco,Spin-bath-Dot,Decoupling-Spinbath,Decoupling-Spinbath2,Decoupling-NV}.

In this work we address this problem by: (i) Enlarging the QPT scenario
for Quantum Ising systems, by generalizing the theory to the case
of a strong high-frequency AC field, and (ii) showing how in principle
this allows the manipulation of the effective Ising Hamiltonian, enabling
one to suppress the spin bath effects. The AC field creates a new
effectively time-independent Hamiltonian for the system, inducing
new interactions and suppressing others, thereby opening up a new
class of QPTs for investigation. By varying the frequency, and intensity
of the field, one also obtains a very rich zero-$T$ phase diagram,
with various new kinds of QCP.

\section{Low energy Hamiltonian}

Well-known solid-state examples of experimental Quantum Ising systems
with long-range dipolar inter-spin interactions include the $LiHo_{x}Y_{1-x}F_{4}$
rare earth system (\cite{aeppli05,RonnowPRB} and \cite{schechter05}-\cite{kycia-dyn}),
and the transition metal-based $Fe_{8}$ molecular spin system \cite{takahashi}.
Recent experiments on 1-dimensional ion trap Quantum Ising chains
(where spin bath effects may be entirely absent) have also successfully
varied the range of the interactions \cite{ionChain1,ionChain2}.
These systems are all described at low energies by Hamiltonians with
spins $\vec{\tau}_{i}$ truncated to the lowest Ising doublet (i.e.,
to its doubly degenerate ground state), separated from the next level
by a gap $\Lambda_{o}\gg|V_{i,j}|,|A|$, where $V_{i,j}$ and $A$
are the strengths of the inter-spin and hyperfine couplings. Then,
their low energy behavior reduces to the study of the next Ising type
Hamiltonian:
\begin{equation}
H\left(t\right)=-\sum_{i=1}^{N}\left[\Delta_{o}+\Gamma\left(t\right)\right]\tau_{i}^{x}-\sum_{i<j}V_{i,j}\tau_{i}^{z}\tau_{j}^{z}+H_{\textrm{HF}}\label{eq:H1}
\end{equation}
The total effective field here is the sum of a constant $\Delta_{o}$
and a time-dependent $\Gamma\left(t\right)$. Typically these are
not real magnetic fields, but effective fields acting in the Hilbert
space of the Ising doublet. Then, it is useful to briefly describe
the truncation procedure: The low-energy effective Hamiltonian is
truncated from a microscopic spin Hamiltonian of the form:
\begin{eqnarray}
H_{M}\left(t\right) & = & -\sum_{i}H_{o}\left(\boldsymbol{S}_{i}\right)+\left[\mathcal{B}_{x}+H_{x}\left(t\right)\right]S_{i}^{x}\label{H-highE}\\
 &  & -\frac{1}{2}\sum_{i,j\neq i}U_{i,j}S_{i}^{z}S_{j}^{z}+H_{\textrm{HF}}^{M}\nonumber 
\end{eqnarray}
where $H_{o}\left(\boldsymbol{S}_{i}\right)$, the local \char`\"{}high-energy\char`\"{}
ionic spin Hamiltonian, acts on spins $\left\{ {\bf S}_{j}\right\} $.
The Hilbert space for each magnetic ion now has dimension $2S+1$.
The high-energy hyperfine coupling takes the form:
\begin{eqnarray}
H_{\textrm{HF}}^{M} & = & \sum_{\mu,\nu}\sum_{j,k}\Lambda_{j,k}^{\mu\nu}S_{j}^{\mu}I_{k}^{\nu}\label{H-hyp-highE}
\end{eqnarray}
where we use a slightly unconventional notation in which $\Lambda_{jk}^{\mu\nu}$
denotes the \char`\"{}bare\char`\"{} hyperfine coupling between the
full spin ${\bf S}_{j}$ and the nuclear spin ${\bf I}_{k}$, and
drop the nuclear quadrupolar couplings, which are negligible for the
quantum Ising systems examined so far. While the low-energy form (Eq.\ref{eq:H1})
is generic to all of the Quantum Ising systems so far investigated,
the high-energy form (Eq.\ref{H-highE}) varies widely from one physical
system to another, depending on the magnitude of the ${\bf S}_{j}$,
the lattice symmetry, the strength of the spin-orbit, crystal field,
hyperfine fields, and so on. Thus the details of the truncation, of
the dependence of the low-energy fields $\Delta_{o}$ and $\Gamma\left(t\right)$
on the external fields, and of the magnitude and anisotropy of the
low-energy $A_{j,k}^{\mu\nu}$ and $V_{i,j}$, depend very much on
which system we are looking at. We will discuss here the 3 cases:

\paragraph{\textbf{$LiHo_{x}Y_{1-x}F_{4}$:}}

In this case, the high-energy Hamiltonian involves ionic spins with
$S=8$ and a local spin Hamiltonian 
\begin{eqnarray}
H_{o}\left({\bf S}\right) & = & \sum_{k=4,6}R_{k}^{4}\left(C\right)\hat{O}_{k}^{4}\left(C\right)+R_{6}^{4}\left(S\right)\hat{O}_{6}^{4}\left(S\right)\label{H-stevens}\\
 &  & +\sum_{k=2,4,6}R_{k}^{0}\hat{O}_{k}^{0}\nonumber 
\end{eqnarray}
written in terms of the standard Stevens operators $\hat{O}_{k}^{q}$
(see, eg., Jensen and MacKintosh \cite{Jensen}); the best values
of the parameters $R_{k}^{q}$ are given by Ronnow et al \cite{aeppli05,RonnowPRB}.
This \char`\"{}high-energy\char`\"{} form is valid up to energy scales
$\sim10^{3}$K. The truncation of the high-energy Hamiltonian of Eqs.\ref{H-highE},\ref{H-stevens}
down to the low-energy form in Eq.\ref{eq:H1} has been thoroughly
discussed in the literature \cite{aeppli05,RonnowPRB,Chack2004,schechter05}.
The low-energy form can be used for energies smaller than the gap
of $\sim11.3$K which exists between the low-energy spin doublet and
a 3rd intermediate state, through which virtual transitions allow
a coupling between the 2 lowest Ising states $|\Uparrow\rangle$ and
$|\Downarrow\rangle$ on each site. The transition matrix element
$\Delta_{o}\left(\mathcal{B}_{x}\right)$ between these states is
a highly non-linear function of $\mathcal{B}_{x}$, obtainable by
exact diagonalization\cite{schechter05}. The hyperfine interactions
are large: experiment shows that $\Lambda_{j,k}^{\mu\nu}\sim0.039$K
for the bare on-site $Ho$ hyperfine coupling, for which $I=7/2$,
with considerably smaller values for the hyperfine couplings to the
four $F$ nuclear spins. This then gives a splitting between adjacent
hyperfine levels of $\sim0.22$K, and a total spread of energy over
the eight hyperfine levels of $\sim1.5$K. Thus the hyperfine energy
scale competes very well with the dipolar coupling $U_{i,j}$ between
nearest neighbor $Ho$ ions, even for the pure $LiHo$ system, where
the energy difference between $|\Uparrow\Uparrow\rangle$ and $|\Uparrow\Downarrow\rangle$
configurations coming from the dipolar interactions is also $\sim1.5$K.
For $Y$-doped $LiHo_{x}Y_{1-x}F_{4}$, this nearest neighbor dipole
coupling is reduced by a factor $\sim O(x)$, and the hyperfine coupling
then dominates.

\paragraph{The $Fe_{8}$ molecular spin system:}

In this case the high-energy Hamiltonian has spin $S=10$ (coming
from a core of eight $Fe$ ions), and a local spin Hamiltonian which
is well approximated by 
\begin{equation}
H_{o}\left({\bf S}\right)=-D_{o}S_{x}^{2}+E_{o}S_{y}^{2}-K_{4}\left(S_{+}^{4}+S_{-}^{4}\right)\label{H-Fe8}
\end{equation}
where the values of $D_{o}$, $E_{o}$, and $K_{4}$ (all positive)
were measured some time ago \cite{barra}. This form can be used up
to energy scales $\sim50$K. Each $Fe_{8}$ molecule has up to 213
different nuclear spins, depending on which of the various $Fe,Br,N,O$,
and $H$ isotopes in the molecule are being used; each of the hyperfine
couplings for this system has been calculated, but they are typically
very small (the proton couplings range from $\sim3$mK down to well
below $1$mK, except for the odd outlier, with similar values for
the other non-metallic nuclear species; the coupling to isotopically
substituted $^{57}Fe$ nuclei is $\sim4$mK). Thus the inter-molecular
dipolar coupling, of a similar magnitude to that for $LiHo$, is much
larger. The truncation to the low-energy doublet form can again be
done numerically \cite{MST06}, and is valid for energies smaller
than the gap of $\sim5$K to the next highest states. Again the dependence
of $\Delta_{o}$ and $\Gamma\left(t\right)$ on a transverse field
$\boldsymbol{\mathcal{B}}$ is very non-linear, and also varies enormously
with the angle of the field in the easy $\hat{x}\hat{y}$-plane -
for $\boldsymbol{\mathcal{B}}$ oriented along the easy $\hat{x}$-axis
one sees very strong oscillations of $\Delta_{o}$ as a function of
$\mathcal{B}_{x}$.

\paragraph{Ionic spin chains:}

In this case one can, to a good approximation, begin by ignoring the
coupling to a spin bath. The Hamiltonian in experiments \cite{ionChain1,ionChain2,ionChain3}
can be more general than the standard Quantum Ising system, and the
low energy form is: 
\begin{eqnarray}
{\cal H}_{eff} & = & \Delta_{o}\sum_{j}\tau_{j}^{x}+\sum_{i\neq j}J_{i,j}^{zz}\tau_{i}^{z}\tau_{j}^{z}+{\cal H}_{J}^{\perp}\label{ionChain}
\end{eqnarray}
where the extra term is just an $XY$ form ${\cal H}_{J}^{\perp}=\sum_{i\neq j}J_{i,j}^{\perp}\left(\tau_{i}^{+}\tau_{j}^{-}+\tau_{i}^{-}\tau_{j}^{+}\right)$.
The parameters $\Delta_{o}$ and $J_{i,j}^{\alpha\alpha}$ are again
effective parameters, related to the original applied fields in ways
described in detail in refs.\cite{ionChain1,ionChain2,ionChain3}.
In different experiments with different ions one can vary these parameters
over a rather wide range; typical values are $10^{-9}~K<J_{o}<10^{-7}~K$,
and $10^{-2}<\Delta_{o}/J_{o}<5$, where $J_{o}$ is a typical nearest
neighbor value for either $J_{i,j}^{zz}$ or $J_{i,j}^{\perp}$. The
interactions typically take a power law form as a function of the
distance $r_{ij}=a_{o}|i-j|$, where $a_{o}$ is the lattice spacing
(typical $2-3\ \mu m$), ie., $|J_{i,j}^{\alpha\alpha}|\sim J_{o}|i-j|^{-p}$,
where in principle one can vary $p$ between $1<p<3$. Provided $\Delta_{o}$
is not too large, the coupling to phonons can be adequately suppressed.
The effective Hamiltonian found in this work applies when we ignore
the \char`\"{}easy-plane\char`\"{} or XY-coupling terms in (\ref{ionChain}),
as the phase diagram with these terms added becomes very rich and
requires a separate study.

\section{Magnus expansion}

In what follows we will work exclusively with the low-energy effective
Hamiltonian given in (\ref{eq:H1}) above, because our topic is the
behavior of a Quantum Ising system in a high-frequency AC field. Thus
we will leave the behavior of the parameters $\Delta_{o}$, $\Gamma\left(t\right)$,
$A_{jk}^{\mu\nu}$, and $V_{ij}$, as functions of the real applied
fields, in the form of undetermined variables in this effective Hamiltonian,
to be determined in practice by a combination of numerical calculation
and experiment on whichever system one is dealing with.

The next step to find a time independent Hamiltonian is to approximate
the full time evolution by a stroboscopic one, in terms of an effective
Hamltonian. For that to be possible, we assume that the frequency
$\omega$ of the time dependent field $\Gamma\left(t\right)$ falls
in the range $\Lambda_{o}>\omega\gg|V_{i,j}|,|A|$. This allows us
to describe the effects of the AC field on the quantum Ising system
by a standard Magnus expansion \cite{Bastidas,Modulated-Ising,TI-QPT,Blanes-Magnus,Magnus-Floquet-Micromotion,Floquet-Magnus-Mananga,Goldman},
in inverse powers of $\omega$. The Magnus expansion is a very powerful
method to extract the stroboscopic time evolution of a time dependent
system, when the frequency of the driving field is large. By means
of a transformation to the interaction picture, ie., $H\left(t\right)\rightarrow\tilde{H}\left(t\right)=U_{1}H\left(t\right)U_{1}^{\dagger}-iU_{1}\dot{U}_{1}^{\dagger}$,
where $U_{1}=e^{i\int dt\sum_{j}\mathbf{\Gamma}\left(t\right)\boldsymbol{\tau}_{j}^{x}}$,
one can also capture the renormalization of parameters produced by
non-perturbative effects of the field. The Magnus expansion then approximates
the time dependent Hamiltonian by a time averaged one given by: 
\begin{eqnarray}
\mathcal{H} & = & \tilde{H}_{0}+\frac{1}{\omega}\sum_{n=1}^{\infty}\frac{1}{n}\left[\tilde{H}_{n},\tilde{H}_{-n}\right]\label{eq:Magnus1}
\end{eqnarray}
where $\tilde{H}_{n}$ is the $n$-th Fourier component of $\tilde{H}\left(t\right)$,
and terms $\sim O\left(\left|A\right|/\omega^{m},\left|V_{i,j}\right|/\omega^{m}\right)$
($m\geq2$) are assumed negligible at high frequency.

In discussing a Magnus expansion, it is important to specify the \char`\"{}initialization
protocol\char`\"{}, ie., the way in which the AC applied field is
ramped up at the beginning. One option often used is to use an \char`\"{}adiabatic
launching protocol\char`\"{} \cite{Goldman}, which consists in reaching
the final non-equilibrium steady state by keeping the system in the
same quasienergy state. Heating can be a problem here, specially due
to the spin-phonon couplings in the system, and the fact that interacting
Floquet systems tend to evolve towards an infinite temperature, featureless
state\cite{Infinite-temperature}. Nevertheless the spin-phonon couplings
are typically rather weak for the Quantum Ising systems being currently
studied; this means that it will take the phonon bath some time to
react to the rapid oscillations of the electronic spins. Thus the
use of pulsed fields is more appropriate, with widely spaced pulses,
as the system then has time for energy relaxation between pulses.
It would also be helpful to have a phonon bath for which the density
of states for undesired Floquet transitions, which could drive the
system out of the steady state, is small; this would stabilize the
Floquet phase described in this work \cite{Population-Floquet}. Finally,
in order to avoid the tendency towards an infinite temperature state,
one could try to control the non-adiabatic corrections during the
adiabatic launching (as if it is ramped too slowly, the system will
reach the infinite temperature state), or combine it with a many-body
localized phase, which would prevent the system from thermalizing\cite{Floquet-MBL}.

\section{Dynamical quantum phase transition}

As a warm up we first consider a 'pure' Quantum Ising QPT, with no
spin bath. We apply a linear AC field $\mathbf{\Gamma}\left(t\right)=\Gamma_{x}\cos\left(\omega t\right)$,
and find that $\left[\tilde{H}_{n},\tilde{H}_{-n}\right]=0$ for all
$n$, leaving only the zeroth Fourier component $\tilde{H}_{0}$ in
Eq. (\ref{eq:Magnus1}) (see Appendix for details), and thus a time-independent
effective Hamiltonian:
\begin{eqnarray}
\mathcal{H}^{o} & = & -\sum_{i}\Delta_{o}\tau_{i}^{x}-\sum_{i,j>i}\left[\tilde{V}_{i,j}^{zz}\tau_{i}^{z}\tau_{j}^{z}+\tilde{V}_{i,j}^{yy}\tau_{i}^{y}\tau_{j}^{y}\right]\label{eq:H-Linear}
\end{eqnarray}
where $\tilde{V}_{i,j}^{zz}\left(\alpha\right)=V_{i,j}\left[1+\mathcal{J}_{0}\left(2\alpha\right)\right]/2$,
$\tilde{V}_{i,j}^{yy}\left(\alpha\right)=V_{i,j}\left[1-\mathcal{J}_{0}\left(2\alpha\right)\right]/2$,
the dimensionless parameter $\alpha=\Gamma_{x}/\omega$ and $\mathcal{J}_{m}\left(\alpha\right)$
is an $m$-th order Bessel function; the superscript in $\mathcal{H}^{o}$
indicates zero hyperfine couplings. Thus the periodic driving modifies
the direction and strength of the inter-spin coupling tensor, and
transforms the Ising model into an XY (strictly a YZ) model with anisotropy
controlled by $\alpha$ (as previously shown by \cite{Bastidas} in
1D). The anisotropic XY model has Ising phase transitions when the
transverse magnetic field $\Delta_{0}=\pm\tilde{V}_{i,j}^{\mu\mu}$,
as well as anisotropic transitions for $\tilde{V}_{ij}^{yy}=\tilde{V}_{ij}^{zz}$,
where the magnetization changes its orientation between $M_{y}^{o}$
and $M_{z}^{o}$ (the super-index indicate the absence of a spin bath
in this section). Note Eq.\ref{eq:H-Linear} is valid in arbitrary
dimension, but the dimensionality plays an important role when calculating
the different statistical averages, as it is discussed next.

To characterize the QPT we must determine the magnetization (ie.,
the order parameter). For that, we calculate the double-time Green's
function\cite{Zubarev} $G_{n,m}^{\alpha,\beta}\left(t,t^{\prime}\right)=-i\theta\left(t-t^{\prime}\right)\langle\left\{ S_{n}^{\alpha}\left(t\right),S_{m}^{\beta}\left(t^{\prime}\right)\right\} \rangle$,
using a $1/Z$ expansion to lowest order ($Z$ is the coordination
number), which coincides with the Random Phase Approximation\cite{RonnowPRB}
(RPA). Under this assumptions the Heisenberg equation of motion for
the Green's function simplifies to:
\begin{eqnarray}
\omega G_{n,m}^{\alpha,\beta} & = & \frac{\langle\left\{ S_{n}^{\alpha},S_{m}^{\beta}\right\} \rangle}{2\pi}+i\sum_{\mu}\epsilon_{\mu\alpha\delta}B_{\mu}G_{n,m}^{\delta,\beta}\\
 &  & +i\sum_{\mu}\epsilon_{\mu\alpha\delta}\left(\sum_{j\neq n}V_{n,j}^{\mu\mu}\langle S_{j}^{\mu}\rangle-\sum_{r}A_{r,n}^{\mu\mu}\langle I_{r}^{\mu}\rangle\right)G_{n,m}^{\delta,\beta}\nonumber 
\end{eqnarray}
and can be generally solved. Note that at this point we are considering
a general magnetic field $B_{\mu}$ and interaction $V_{n,j}^{\mu\mu}$,
as it does not complicate things. The last step is to relate the Green's
function with the magnetization using:
\begin{equation}
\langle S_{m}^{\beta}S_{n}^{\alpha}\rangle=i\int\frac{G_{n,m}^{\alpha,\beta}\left(\omega+i\epsilon\right)-G_{n,m}^{\alpha,\beta}\left(\omega-i\epsilon\right)}{e^{\beta\omega}+1}d\omega
\end{equation}
and derive the ``self consistency equation'' (SCE) for the magnetization,
which in absence of the spin bath, simply is:
\begin{eqnarray}
M_{\mu}^{o} & = & \frac{B_{\mu}+M_{\mu}^{o}\tilde{V}_{0}^{\mu\mu}}{2\tilde{\omega}_{S}}\tanh\left(\frac{\beta\tilde{\omega}_{S}}{2}\right)\label{eq:Self-consistency0}
\end{eqnarray}
where $M_{\mu}^{o}$ ($\mu=x,y,z$) is the magnetization along the
$\mu$-axis, $\tilde{\omega}_{S}=\sqrt{\sum_{\mu}\left(B_{\mu}+M_{\mu}^{o}\tilde{V}_{0}^{\mu\mu}\right)^{2}}$,
the effective field is $B_{\mu}=\left(\Delta_{o},0,0\right)$, and
$\tilde{V}_{0}^{\mu\mu}$ is the $\mathbf{q}=0$ Fourier component
of the effective spin-spin interaction along the $\mu$-axis. The
detailed derivation is delegated to the Appendix, as it closely follows
the one of ref.\cite{Fermion-GF}, with the addition of the spin bath.

Note that the formalism applied here is valid for arbitrary large
spins (which is important to describe the $7/2$ spin bath in the
next section). Also the $1/Z$ scaling implies that the results are
expected to be more accurate in higher dimensional systems (where
the coordination number is large), such as the 3D Ising model. Therefore
our results for the magnetization will better describe the experiments
on $LiHo_{x}Y_{1-x}F_{4}$, than the ones on ionic spin chains, where
one would expect large quantum fluctuations which would modify the
magnetization. The system's dimension is encoded in the coordination
number $Z$, or equivalently in the zeroth Fourier component of the
interaction potential $V_{0}$. For the simplest case of nearest neighbor
interaction, one finds that they are related by $V_{0}=ZV$, however
in more general situations, as for the case of long-range dipolar
interactions, this relationship fails and it is more convenient to
just estimate $V_{0}$ by other means, keeping in mind that the system's
dimension is somehow encoded in this value. The inverse temperature
$\beta$ in Eq.\ref{eq:Self-consistency0} corresponds to the phonon
bath temperature that in general, would couple to the spin system.
Although its definition is not generally possible in the presence
of a driving field, we will discuss in the last section how, under
some circumstances, one can still make use of it.

\begin{figure}
\includegraphics[scale=0.55]{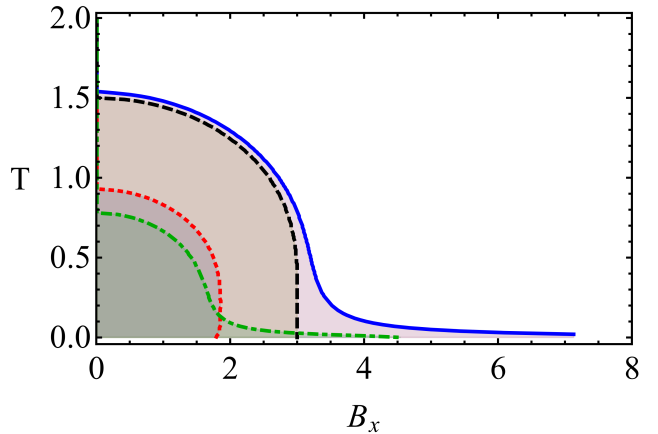}

\caption{\label{fig:Phase-diagram1}Phase diagram of the AC driven transverse
Ising model vs transverse field $B_{x}$ and temperature. The dashed
black line corresponds to the undriven system in absence of hyperfine
coupling. The blue line corresponds to the undriven system coupled
to the $I=7/2$ spin bath. The red line corresponds to the AC driven
system coupled to the spin bath for $\Gamma_{x}/\omega\sim2.4$ (i.e.,
$\mathcal{J}_{0}\left(\alpha\right)=0$). The green line corresponds
to the case $2\Gamma_{x}/\omega\sim2.4$, where the spin-spin interaction
is symmetrical and resembles the Ising system again. For the plots
we used: $V_{0}^{zz}=6\textrm{ K}$ and $A_{0}^{z,\left(\perp\right)}=0.2\left(0.02\right)\textrm{K}$.
These values are chosen so that the phase diagram agrees with the
experimental data in \cite{Chack2004}.}
\end{figure}

Eq.\ref{eq:Self-consistency0} allows to easily compare the AC driven
and the undriven case. For the undriven ($\Gamma_{x}=0$) pure 3D
Ising model we find the next ground state magnetization:
\begin{itemize}
\item For $\Delta_{o}<V_{0}^{zz}/2$: 
\begin{equation}
M_{x}^{0}=\frac{\Delta_{o}}{V_{0}^{zz}},\quad M_{y}^{0}=0,\quad M_{z}^{0}=\pm\frac{\sqrt{\left(V_{0}^{zz}\right)^{2}-4\Delta_{o}^{2}}}{2V_{0}^{zz}}.
\end{equation}
\item For $\Delta_{o}>V_{0}^{zz}/2$: 
\begin{equation}
M_{x}^{0}=\frac{1}{2},\quad M_{y,z}^{0}=0.
\end{equation}
\end{itemize}
The critical field is given by $B_{c}=V_{0}^{zz}/2$, and the finite-$T$
solution gives a zero field Curie temperature $T_{c}=V_{0}^{zz}/4$.
For the case of long-range dipolar interactions, present in $LiHo_{x}Y_{1-x}F_{4}$,
one can directly estimate the zeroth Fourier component $V_{0}^{zz}$
by numerical means\cite{RonnowPRB}, or as we have done in our case,
extract its value from experimental measurements\cite{Chack2004}.
The phase diagram is shown in Fig.\ref{fig:Phase-diagram1}, where
the dashed black line separates a spin ordered FM phase for $\Delta_{o}<B_{c}$
and a paramagnetic one otherwise.

In the presence of the AC field the system is described by the anisotropic
XY model (Eq.\ref{eq:H-Linear}). The anisotropy factor $\tilde{V}_{j,l}^{zz}-\tilde{V}_{j,l}^{yy}=\mathcal{J}_{0}\left(2\alpha\right)V_{j,l}$
becomes a function of $\alpha$ and the Ising model is recovered when
$\mathcal{J}_{0}\left(2\alpha\right)=0$. This means that anisotropic
quantum phase transitions happen every time $\mathcal{J}_{0}\left(2\alpha\right)=0$,
and the magnetization changes its direction between $M_{z}^{o}$ and
$M_{y}^{o}$ for temperatures below a critical $T_{c}$. Ising transitions
could also be induced when $\alpha$ is tuned, as the critical field
$B_{c}$ oscillates between a maximum and a minimum value (red solid
line in Fig.\ref{fig:CriticalField}). Therefore, if the DC field
is within this window, one would also observe PM/FM transitions.

As a remark, it is important to ask whether the high frequency corrections
of order $\omega^{-2}$ in Eq.\ref{eq:Magnus1}, can change the previous
results in a relevant way. The reason is that although their contribution
seems to be small for a high frequency field, we are considering a
time dependent system, where initially small contributions can grow
over time considerably. We will devote this discussion to the next
section, where the spin bath is included.

\section{Nuclear spin bath effects}

In Quantum Ising systems the spin bath effects are often dominated
by a single species of nuclear spin $I_{r}^{\mu}$ at positions ${\bf r}_{r}$.
Let us assume an effective hyperfine coupling in Eq.\ref{eq:H1} given
by:
\begin{equation}
H_{\text{HF}}=\sum_{\mu,r,j}A_{r,j}^{\mu}I_{r}^{\mu}S_{j}^{\mu}\label{Eq:H-Hyp}
\end{equation}
with principal axes along $\mu=x,y,z$ (the generalization to more
complex forms is straightforward). Two spin bath mechanisms can then
strongly affect Quantum Ising systems:

(i) \textit{Transverse blocking mechanism}: In a quantum Ising system
with hyperfine coupling, the electronic spin cannot simply flip between
$|\Uparrow\rangle$ and $|\Downarrow\rangle$; it must carry the nuclear
spin with it. However, transitions between $|\Downarrow\uparrow\rangle$
and $|\Uparrow\downarrow\rangle$ can then no longer be mediated by
$\Delta_{o}$, which do not flip the nuclear spin. Transverse hyperfine
interactions could produce this flip, but in many real Quantum Ising
systems, the effective longitudinal hyperfine coupling $A_{0}^{z}$
is often much stronger than the transverse one (the large $g$ factor
anisotropy of any Ising system also forces a strong anisotropy in
the Ising hyperfine coupling). The spin bath then strongly suppresses
transverse electronic spin fluctuations \cite{schechter05}, and the
system reverts to classical Ising behavior until $\Delta_{o}$ is
large enough to overcome $A_{0}^{z}$. This changes the phase diagram,
shifting the critical point towards large values of $\Delta_{0}$(see
Fig.\ref{fig:Phase-diagram1}, blue line), and it also radically alters
the electronic spin dynamics. Many features of the resulting experimental
behavior (such as the gapping of the electronic exciton mode in $LiHo$,
even at the QCP \cite{aeppli05}), are still not properly understood.

(ii) \textit{Spin bath decoherence}: The spin bath causes decoherence
in the electronic spin dynamics \cite{spinB,takahashi}. Such decoherence
blocks the use of Quantum Ising systems as quantum information processors,
for which they are otherwise ideally suited.

It would clearly be desirable to control the strength of both the
inter-spin and the hyperfine couplings, and if possible, to suppress
the hyperfine coupling completely. As we now see, this can be done
in strong AC fields. We treat the hyperfine coupling $H_{\text{HF}}=\sum_{r,j,\mu}A_{r,j}^{\mu}I_{r}^{\mu}S_{j}^{\mu}$
in an AC field using the same maneuvers as above; $H_{\text{HF}}$
is then renormalized to: 
\begin{equation}
\mathcal{H}_{\text{HF}}=\sum_{\mu,j,r}\tilde{A}_{r,j}^{\mu}I_{r}^{\mu}S_{j}^{\mu}\label{eq:HyperfineCoupling}
\end{equation}
where the effective coupling is $\tilde{A}_{r,j}^{\mu}=\left(A_{r,j}^{x},A_{r,j}^{y}\mathcal{J}_{0}\left(\alpha\right),A_{r,j}^{z}\mathcal{J}_{0}\left(\alpha\right)\right)$.
Because the field couples to the electronic spins, but not to the
nuclear spins (the nuclear Zeeman coupling $\ll|A_{r,j}^{\mu}|$),
the renormalization of $A_{r,j}^{\mu}$ is different from that of
$V_{i,j}$ (with factors $\mathcal{J}_{0}\left(\alpha\right)$ rather
than $\left(1\pm\mathcal{J}_{0}\left(2\alpha\right)\right)/2$, and
with $\alpha$ rather than $2\alpha$ in the argument). Thus, by tuning
the AC field amplitude, we can either (i) tune the inter-spin interaction
$V_{i,j}$, to study the spin bath effects, or (ii) suppress the longitudinal
hyperfine coupling $A_{z}$, to study the effects of $V_{i,j}$ in
isolation.

To quantify all of this, we make use of the self-consistent equations
for the magnetization, now including the hyperfine coupling to the
nuclear spin bath (to be specific this is done for the case $I=7/2$,
appropriate to the $LiHo$ system). We then arrive at the pair of
equations for the magnetization of the electronic system and nuclear
bath ($M_{\mu}$ and $m_{\mu}$, respectively): 
\begin{eqnarray}
M_{\mu} & = & \frac{B_{\mu}+M_{\mu}\tilde{V}_{0}^{\mu\mu}-m_{\mu}\tilde{A}_{0}^{\mu}}{2\tilde{\omega}_{S}}\tanh\left(\frac{\beta\tilde{\omega}_{S}}{2}\right)\label{eq:M-Total}\\
m_{\mu} & = & -\frac{\tilde{A}_{0}^{\mu}M_{\mu}}{2\tilde{\omega}_{B}}\left[\tanh\left(\frac{\beta\tilde{\omega}_{B}}{2}\right)\right.\\
 &  & \left.+2\tanh\left(\beta\tilde{\omega}_{B}\right)+4\tanh\left(2\beta\tilde{\omega}_{B}\right)\right]\nonumber 
\end{eqnarray}
where $\tilde{\omega}_{S}=\sqrt{\sum_{\mu}\left(B_{\mu}+M_{\mu}\tilde{V}_{0}^{\mu\mu}-m_{\mu}\tilde{A}_{0}^{\mu}\right)^{2}}$
is the system quasiparticle spectrum, and $\tilde{\omega}_{B}=\sqrt{\sum_{\mu}\left(M_{\mu}\tilde{A}_{0}^{\mu}\right)^{2}}$
is the spin bath quasiparticle spectrum. In order to show the ``transverse
blocking mechanism'', in the Appendix we approximate the equation
for $M_{z}$ when $A_{0}^{z}\gg A_{0}^{x,y}$ and find that $M_{z}\simeq\frac{M_{z}V_{0}^{z}}{2B_{x}}+\frac{7A_{0}^{z}}{4B_{x}}$;
this indicates that $M_{z}=0$ is not a solution due to a remnant
magnetization proportional to $A_{0}^{z}/B_{x}$.

Setting the field amplitude to $\mathcal{J}_{0}\left(\alpha\right)=0$,
one can see that the longitudinal hyperfine coupling $\tilde{A}_{0}^{z,y}$
vanishes, and only the transverse part $A_{0}^{x}$ remains. Then,
as the hyperfine interaction only acts in the longitudinal direction,
one finds a renormalization of the critical field $B_{c}$ to smaller
values, but the QPT is still well defined, as it is driven by the
transverse field $\Delta_{0}$(Fig.\ref{fig:Phase-diagram1}, dotted
red). Furthermore, as for this value of $\alpha$ one has $\tilde{V}_{0}^{yy}>\tilde{V}_{0}^{zz}$,
the ferromagnetic phase is now magnetized along the $y$ axis.

\begin{figure}
\includegraphics[scale=0.55]{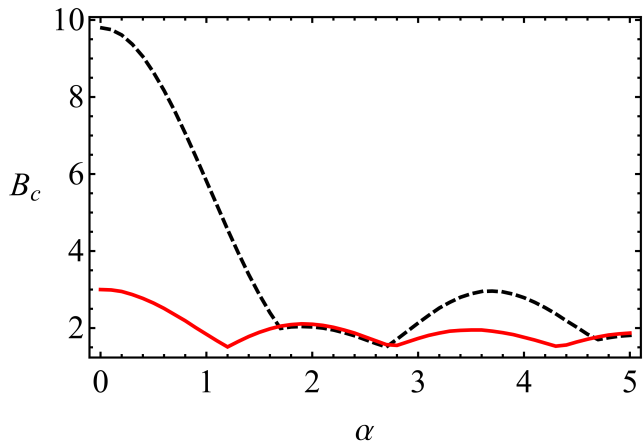}

\caption{\label{fig:CriticalField}The dashed black line shows the critical
field $B_{c}$ as a function of the ratio $\alpha=\Gamma_{x}/\omega$
for the AC driven Ising system coupled to a $7/2$ spin bath. The
red solid line corresponds to the AC driven system in absence of the
spin bath. At low amplitude, the spin bath contributes very strongly
and drastically changes the critical field due to the blocking mechanism.
As the amplitude increases, the effect of the bath is removed and
only the transverse part $A_{0}^{x}$ contributes with a small shift.
There are some regions where $B_{c}$ in presence of a bath is even
lower than in the absence of hyperfine coupling ($\alpha\sim1.7-2.7$).
These are the regions where $\tilde{A}_{0}^{y,z}$ changes sign and
overcomes $A_{0}^{x}$. The parameters are fixed according to the
experimental ones for $LiHoF_{4}$:$V_{0}^{zz}=6$ K, $A_{0}^{z}=0.2$
K and $A_{0}^{x,y}=0.02$ K.}
\end{figure}

Similarly, one can choose other values of $\alpha$ such as $\mathcal{J}_{0}\left(2\alpha\right)=0$,
where the asymmetry factor vanishes, and then the Ising model is recovered,
or one can tune the sign of $\tilde{A}_{0}^{z,y}$, changing the ground
state properties to a triplet state $\left\{ |\Uparrow\uparrow\rangle,|\Downarrow\downarrow\rangle\right\} $.
In Fig.\ref{fig:CriticalField} we plot the critical field $B_{c}$
for the AC driven Ising model coupled to the spin bath, as a function
of the AC field parameter $\alpha$, for $T=0$.

This plot shows that for small $\alpha$, the system behaves as in
the undriven case, where the spin bath greatly affects the value of
the critical field due to the blocking mechanism. As this blocking
is produced by the difference between the transverse $A_{0}^{x}$
and the longitudinal hyperfine coupling $A_{0}^{y,z}$, and the later
is renormalized by the AC field, one can observe that by increasing
$\alpha$ the system approaches the isolated system behavior. Therefore
it would be possible to experimentally analyze the opposite regimes
of ideal Ising QPT in absence and in presence of a spin bath by just
tuning the external AC field.

As we previously pointed out, it is important to discuss the effect
of the high frequency corrections neglected in the Magnus expansion
(Eq.\ref{eq:Magnus1}). We have calculated the next order leading
term $\frac{1}{2\omega^{2}}\sum_{n=1}^{\infty}\frac{1}{n^{2}}\left(\left[\left[\tilde{H}_{n},\tilde{H}_{0}\right],\tilde{H}_{-n}\right]+h.c.\right)$,
and although the effective Hamiltonian contains now up to four-body
interactions, they are all weighted by Bessel functions and a factor
$\omega^{-2}$, which in general give corrections one or two orders
of magnitude smaller than $\tilde{H}_{0}$. Importantly, we find that
the transverse blocking mechanism, produced due to the initially large
anisotropy between $A_{0}^{z}$ and $A_{0}^{x,y}$, is not restored
by the second order corrections, and the renormalized critical point
should not be greatly affected. Nevertheless one should be careful
with the growth of high frequency corrections for large times; this
would restrict the maximum duration of the experiments to times shorter
than the inverse of the energy correction. In addition, the time control
can be complex due to the competition between the initialization time
and the infinite temperature limit of interacting Floquet systems\cite{Infinite-temperature},
but several strategies based on the properties of the transient dynamics
could allow to overcome this issue\cite{Thermalization,Floquet-MBL}.
As a check we have included in the Appendix the simulation of the
dynamics of the Quantum Ising system coupled to a spin bath, when
the QCP is crossed from the ferromagnetic phase. It shows that in
absence of the AC field, the QPT to the paramagnetic phase is suppressed
due to the spin bath, but in the presence of the AC field tuned to
$\mathcal{J}_{0}\left(\alpha\right)=0$, the time average magnetization
$\int_{0}^{T}M_{z}\left(t\right)dt$ vanishes, indicating the cancellation
of the longitudinal hyperfine coupling $\tilde{\mathcal{A}}_{0}^{z}=0$.
Therefore, one can conclude that the high frequency corrections do
not affect the suppression of the hyperfine interaction, at least
within the time scales of the simulation.

\section{Concluding remarks}

We have obtained a static effective Hamiltonian for the AC driven
transverse Ising model in the presence of a spin bath, in which the
inter-spin and the hyperfine interactions are renormalized as a function
of the AC field intensity and frequency. We have found that the inter-spin
and hyperfine interaction renormalize differently, which allows to
study the Ising QPT in a large number of cases ranging from positive
to negative hyperfine interaction (Fig.\ref{fig:CriticalField}).
The effective Hamiltonian for the AC driven Quantum Ising model in
Eq.\ref{eq:H-Linear}, and the effective hyperfine interaction in
Eq.\ref{eq:HyperfineCoupling} are general results, valid in arbitrary
dimension; however the magnetization in Eq.\ref{eq:M-Total} is calculated
to lowest order and should be more accurate for higher dimensional
systems such as the 3D Quantum Ising model. Importantly, the equations
for the magnetization in the presence of the spin bath are derived
for arbitrary large spins, which allows to apply this theory to different
types of spin bath. Finally, the phase diagram in Eq.\ref{fig:Phase-diagram1}
is obtained as a function of the transverse field and the temperature
$T=1/\beta$, which is in general ill-defined for non-equilibrium
situations such as the one with the AC field. The temperature is set
by a phonon bath, and in order to avoid heating due to the AC field,
a pulsed field experiment would be useful, with the pulses short enough
so that the spin-phonon couplings have no time to heat the phonon
bath. As we previously discussed, the initialization protocol should
be engineered so as to reach the desired steady state, which can be
done using adiabatic launching. Furthermore, in some cases the interactions
with the phonon bath could be used to stabilize the steady state\cite{Population-Floquet}

The classic QPT magnetic insulator $LiHo_{x}Y_{1-x}F_{4}$, with spin
$S=8$ magnetic ions, perhaps the canonical Quantum Ising system,
displays quantum annealing \cite{Qanneal} and a quantum spin glass
phase \cite{QSpGl}, as well as quantum critical behavior \cite{Qcrit}.
However, the strong coupling to the nuclear spin bath disrupts completely
the expected quantum critical behavior around the QCP \cite{aeppli05,schechter05}
(and leads to various other dynamic and thermodynamic effects \cite{giraud,luke10,kycia-dyn}).
For nearest neighbor spins, $|V_{i,j}|\rightarrow V\sim1.2x$, in
K units, and the hyperfine level splitting $|A|\sim0.22$K (with spin
$I=7/2$). In a high frequency AC field ($\Lambda_{o}\gg\hbar\omega\gg|V_{i,j}|,|A_{r,j}^{\mu}|$,
ie, for $\omega\sim30-200$ GHz), we may then directly apply the theory
given here. The results are shown in Figs. \ref{fig:Phase-diagram1}
and \ref{fig:CriticalField}), and these constitute predictions for
this system.

In the $Fe_{8}$ system the hyperfine couplings are much smaller,
and can be varied by isotopic substitution\cite{WW00}. Because there
is a whole spectrum of these couplings, one cannot suppress them all
simultaneously - but one can select out particular groups of nuclear
spins for suppression, and the effective couplings as a function of
applied field are well understood \cite{MST06,takahashi}. What is
interesting here is the possibility of controlling the longitudinal
dipolar coupling between molecules, allowing one to look at single
molecule dynamics.

In ion trap spin chains we can typically discount spin bath effects.
What is interesting here is the possibility of varying the range of
the inter-spin interactions, as well as their strength, and observing
the spin dynamics in real time for both short- and long-range interaction
forms \cite{ionChain2}; calculations including corrections to the
RPA result are underway to give quantitative predictions, as they
can be important for low dimensional systems. In these systems one
can also introduce transverse inter-spin interactions - this makes
the eventual phase diagram very rich indeed.

In all three systems experimental testing of the results herein should
be easily possible - quantitative comparison will require numerical
work, and we emphasize that in real experiments one will need to take
account of demagnetization fields, which are in general inhomogeneous
in real systems. This will need to be evaluated numerically (compare
\cite{takahashi}), and experiments with ``whisker''-shaped samples
(for solid-state systems) would be useful.

This work has been supported by NSERC of Canada, and by PITP. We acknowledge
helpful discussions with G. Aeppli on the experimental constraints.\begin{widetext}

\appendix

\section{Magnus expansion}

To quantify the renormalization effects produced by a large amplitude
of the AC field, one must first use a transformation to the interaction
picture:
\begin{equation}
\tilde{H}\left(t\right)=U_{1}H\left(t\right)U_{1}^{\dagger}-iU_{1}\dot{U}_{1}^{\dagger}=-\sum_{i}\Delta_{o}\tilde{\tau}_{i}^{x}-\frac{1}{2}\sum_{i,j\neq i}V_{i,j}\tilde{\tau}_{i}^{z}\tilde{\tau}_{j}^{z}+\sum_{\mu,r,j}A_{r,j}^{\mu}I_{r}^{\mu}\tilde{\tau}_{j}^{\mu}
\end{equation}
where $U_{1}=\exp\left[i\int\sum_{j}\Gamma\left(t\right)\tau_{j}^{x}\right]dt$,
with $\mu=x,y,z$ and $\tilde{\tau}_{j}^{\mu}\equiv U_{1}\tau_{j}^{\mu}U_{1}^{\dagger}$.
For the calculation of the Magnus expansion one needs the Fourier
coefficients of the time dependent Hamiltonian. They are given by
the next expressions: 
\begin{equation}
\tilde{H}_{0}=-\sum_{i}\Delta_{o}\tau_{i}^{x}-\sum_{i,j\neq i}\frac{V_{i,j}}{4}\left\{ \left[1+\mathcal{J}_{0}\left(2\alpha\right)\right]\tau_{i}^{z}\tau_{j}^{z}+\left[1-\mathcal{J}_{0}\left(2\alpha\right)\right]\tau_{i}^{y}\tau_{j}^{y}\right\} +\sum_{\mu,r,j}\tilde{A}_{r,j}^{\mu}I_{r}^{\mu}\tau_{j}^{\mu}
\end{equation}
\begin{eqnarray}
\tilde{H}_{\pm\left(2n+1\right)} & = & \pm i\sum_{i,j\neq i}\frac{V_{ij}\mathcal{J}_{2n+1}\left(2\alpha\right)}{4}\left(\tau_{i}^{y}\tau_{j}^{z}+\tau_{i}^{z}\tau_{j}^{y}\right)\pm i\mathcal{J}_{2n+1}\left(\alpha\right)\sum_{r,j}\left(A_{r,j}^{y}I_{r}^{y}\tau_{j}^{z}-A_{r,j}^{z}I_{r}^{z}\tau_{j}^{y}\right)\\
\tilde{H}_{\pm2\left(n+1\right)} & = & -\sum_{i,j\neq i}\frac{V_{ij}\mathcal{J}_{2n+2}\left(2\alpha\right)}{4}\left(\tau_{i}^{z}\tau_{j}^{z}-S_{i}^{y}\tau_{j}^{y}\right)+\mathcal{J}_{2\left(n+1\right)}\left(\alpha\right)\sum_{r,j}\left(A_{r,j}^{y}I_{r}^{y}\tau_{j}^{y}+A_{r,j}^{z}I_{r}^{z}\tau_{j}^{z}\right)
\end{eqnarray}
where $\alpha\equiv\Gamma_{x}/\omega$, $\mathcal{J}_{m}\left(\alpha\right)$
is an $m$-th order Bessel function, and where the renormalized hyperfine
couplings are 
\begin{equation}
\tilde{A}_{r,j}^{x}=A_{r,j}^{x},\quad\tilde{A}_{r,j}^{y}=A_{r,j}^{y}\mathcal{J}_{0}\left(\alpha\right),\quad\tilde{A}_{r,j}^{z}=A_{r,j}^{z}\mathcal{J}_{0}\left(\alpha\right).\label{R-cplg-A}
\end{equation}
From these expression we see that $\left[\tilde{H}_{n},\tilde{H}_{-n}\right]=0$,
and to order $1/\omega$ we only need to use $\tilde{H}_{0}$ in the
expansion. The effective time-independent Hamiltonian becomes: 
\begin{eqnarray}
\mathcal{H} & = & -\sum_{i}\Delta_{o}\tau_{i}^{x}-\frac{1}{2}\sum_{i,j\neq i}^{N}\left[\tilde{V}_{ij}^{zz}\tau_{i}^{z}\tau_{j}^{z}+\tilde{V}_{ij}^{yy}\tau_{i}^{y}\tau_{j}^{y}\right]+\sum_{\mu,r,j}\tilde{A}_{r,j}^{\mu}I_{r}^{\mu}\tau_{j}^{\mu}\label{eq:H-Linear-1}
\end{eqnarray}
in which the renormalized couplings are:
\begin{eqnarray}
\tilde{V}_{i,j}^{zz}\left(\alpha\right) & = & V_{i,j}\left[1+\mathcal{J}_{0}\left(2\alpha\right)\right]/2\label{R-cplg}\\
\tilde{V}_{ij}^{yy}\left(\alpha\right) & = & V_{i,j}\left[1-\mathcal{J}_{0}\left(2\alpha\right)\right]/2
\end{eqnarray}
We see that under the effect of the AC field the model becomes an
effective anisotropic \char`\"{}$XY$\char`\"{} (actually, $YZ$)
spin system in a transverse field. Thus one effect of the AC field
is to modify the very strong dominance of the \char`\"{}zz\char`\"{}
coupling in the effective dipolar interaction between Ising spins.
We note also that the argument of the Bessel functions in the renormalized
hyperfine couplings is half that involved in the renormalized inter-spin
couplings.

\section{Magnetization calculation}

Here we include the details of the calculation for the magnetization
self-consistency equations in presence of the spin bath. A more general
form of the Hamiltonian discussed in this paper is: 
\begin{equation}
H=-\sum_{\mu,j}B_{\mu}S_{j}^{\mu}-\frac{1}{2}\sum_{\mu}\sum_{j,l\neq j}V_{j,l}^{\mu}S_{j}^{\mu}S_{l}^{\mu}+\sum_{\mu,r,j}A_{r,j}^{\mu}I_{r}^{\mu}S_{j}^{\mu}
\end{equation}
where $\alpha,\beta=x,y,z$, and $S_{n}^{\alpha}$ operates on an
arbitrary spin of spin S (not just $S=1/2$) at site $n$; the nuclear
spin $I_{r}^{\mu}$ also takes arbitrary value. We are interested
in the Green's function for the calculation of the magnetization,
defined by: 
\begin{equation}
G_{n,m}^{\alpha,\beta}\left(t,t^{\prime}\right)=-i\langle S_{n}^{\alpha}\left(t\right);S_{m}^{\beta}\left(t^{\prime}\right)\rangle
\end{equation}
where $\langle\ldots\rangle$ corresponds to the statistical average
with respect to the thermal density matrix $\rho=e^{-\beta H}$, and
the semi-colon indicates that we can consider the time ordered, retarded
or advanced Green's functions (they all have the same equation of
motion). The corresponding equation of motion for the electronic spins
is given by: 
\begin{align}
\omega G_{n,m}^{\alpha,\beta}\left(\omega\right) & =\frac{1}{2\pi}\langle\left\{ S_{n}^{\alpha},S_{m}^{\beta}\right\} \rangle+i\sum_{\mu}\epsilon_{\mu\alpha\delta}B_{\mu}G_{n,m}^{\delta,\beta}\left(\omega\right)\\
 & +i\sum_{\mu}\epsilon_{\mu\alpha\delta}\sum_{j\neq n}V_{n,j}^{\mu}G_{jn,m}^{\mu\delta,\beta}\left(\omega\right)-i\sum_{\mu,r}\epsilon_{\mu\alpha\delta}A_{r,n}^{\mu}K_{rn,m}^{\mu\delta,\beta}\left(\omega\right)\nonumber 
\end{align}
where we have defined: 
\begin{equation}
G_{jn,m}^{\mu\delta,\beta}\left(t,t^{\prime}\right)=-i\langle S_{j}^{\mu}\left(t\right)S_{n}^{\delta}\left(t\right);S_{m}^{\beta}\left(t^{\prime}\right)\rangle,\quad K_{rn,m}^{\mu\delta,\beta}\left(t,t^{\prime}\right)=-i\langle I_{r}^{\mu}\left(t\right)S_{n}^{\delta}\left(t\right);S_{m}^{\beta}\left(t^{\prime}\right)\rangle
\end{equation}
The expression above is valid for arbitrary spin values. Note that
we have used anti-commutation relationships for the definition of
the Green's functions, as it is more convenient for the underlying
pole structure that we will encounter later on. In what follows we
adapt the spin operator decoupling methods discussed by, eg., Wang
et al. \cite{Fermion-GF}, for lattice electronic spins, to the more
general case of a set of lattice spins coupled to nuclear spins.

We decouple the higher Green functions in the equation of motion neglecting
correlations between different sites. This approximation can be understood
from the perspective of a $1/Z$ expansion, being $Z$ the coordination
number of the system. It is known that to lowest order, ie neglecting
quantum correlations, the $1/Z$ expansion agrees with the Random
Phase Approximation (RPA) and that for higher dimensional systems
such as the 3D Ising model considered for $LiHoF$, it should provide
reasonable good results\cite{RonnowPRB,aeppli05}. Once applied the
decoupling scheme, we find: 
\begin{align}
\omega G_{n,n}^{\alpha,\beta} & =\frac{\chi_{\alpha,\beta}}{2\pi}+i\sum_{\mu}\epsilon_{\mu\alpha\delta}\left(B_{\mu}+\sum_{j\neq n}V_{n,j}^{\mu}\langle S_{j}^{\mu}\rangle-\sum_{r}A_{r,n}^{\mu}\langle I_{r}^{\mu}\rangle\right)G_{n,n}^{\delta,\beta}
\end{align}
where $\chi_{\alpha\beta}=\langle\left\{ S_{n}^{\alpha},S_{n}^{\beta}\right\} \rangle$
are the spin anti-commutators (we suppress the site index $n$) and
$G_{n,n}^{\alpha,\beta}$ is now the Green's function in the RPA approximation.
The calculation of the Green's functions is fairly straightforward
\cite{Fermion-GF}); we rewrite the system of equations as: 
\begin{equation}
\left(\boldsymbol{1}\omega-\boldsymbol{H}\right)\boldsymbol{G}\left(\omega\right)=\boldsymbol{F},\ \boldsymbol{G}\left(\omega\right)=\left(\begin{array}{c}
G_{n,n}^{x,\beta}\left(\omega\right)\\
G_{n,n}^{y,\beta}\left(\omega\right)\\
G_{n,n}^{z,\beta}\left(\omega\right)
\end{array}\right)
\end{equation}
where 
\begin{equation}
\boldsymbol{F}=\frac{1}{2\pi}\left(\begin{array}{c}
\chi_{x,\beta}\\
\chi_{y,\beta}\\
\chi_{z,\beta}
\end{array}\right),\ \boldsymbol{H}=\left(\begin{array}{ccc}
0 & iH_{z} & -iH_{y}\\
-iH_{z} & 0 & iH_{x}\\
iH_{y} & -iH_{x} & 0
\end{array}\right),
\end{equation}
and $H_{\alpha}$ are the components of an effective field defined
as 
\begin{equation}
H_{\alpha}=B_{\alpha}+M_{\alpha}V_{0}^{\alpha}-m_{\alpha}A_{\alpha},
\end{equation}
$M_{\alpha}$ is the electronic spin magnetization, and $m_{\alpha}$
the nuclear spin bath magnetization. The $\boldsymbol{H}$ matrix
can be diagonalized and has eigenvalues $\omega=\left\{ 0,\pm\sqrt{\sum_{\alpha}H_{\alpha}^{2}}\right\} $.

The Green's functions can be obtained as: 
\begin{equation}
G_{n,n}^{\alpha,\beta}=\sum_{\lambda=1}^{3}\sum_{\tau=1}^{3}\frac{U_{\alpha\tau}U_{\tau\lambda}^{-1}}{\omega-\omega_{\tau}}F^{\lambda,\beta}=\sum_{\lambda=1}^{3}R^{\alpha,\lambda}F^{\lambda,\beta}
\end{equation}
where $U$ is the matrix that diagonalizes $\boldsymbol{H}$. From
this expression we calculate the statistical averages straightforwardly
using: 
\begin{equation}
\langle S_{n}^{\beta}S_{n}^{\alpha}\rangle=i\int\frac{G_{n,n}^{\alpha,\beta}\left(\omega+i\epsilon\right)-G_{n,n}^{\alpha,\beta}\left(\omega-i\epsilon\right)}{e^{\beta\omega}+1}d\omega
\end{equation}
\begin{equation}
\langle S_{n}^{\beta}S_{n}^{\alpha}\rangle=\sum_{\lambda=1}^{3}\sum_{\tau=1}^{3}\frac{U_{\alpha\tau}U_{\tau\lambda}^{-1}}{e^{\beta\omega_{\tau}}+1}\tilde{F}^{\lambda,\beta}
\end{equation}
where 
\begin{equation}
\tilde{F}^{\lambda\beta}=2\pi F^{\lambda\beta}=\chi^{\lambda\beta}
\end{equation}
In order to simplify the expression we can use the relation between
commutators and anti-commutators: 
\begin{eqnarray}
\langle\left\{ S_{n}^{\alpha},S_{n}^{\beta}\right\} \rangle & = & \langle\left[S_{n}^{\alpha},S_{n}^{\beta}\right]\rangle+2\langle S_{n}^{\beta}S_{n}^{\alpha}\rangle
\end{eqnarray}
Writing $\Gamma=\sum_{\tau=1}^{3}\frac{U_{\alpha\tau}U_{\tau\lambda}^{-1}}{e^{\beta\omega_{\tau}}+1}$,
the matrix equation for the statistical averages becomes:
\begin{eqnarray}
\left(\boldsymbol{1}-2\boldsymbol{\Gamma}\right)\left(\begin{array}{c}
\langle S_{n}^{\beta}S_{n}^{x}\rangle\\
\langle S_{n}^{\beta}S_{n}^{y}\rangle\\
\langle S_{n}^{\beta}S_{n}^{z}\rangle
\end{array}\right) & = & \boldsymbol{\Gamma}\left(\begin{array}{c}
\langle\left[S_{n}^{\alpha},S_{n}^{\beta}\right]\rangle\\
\langle\left[S_{n}^{\alpha},S_{n}^{\beta}\right]\rangle\\
\langle\left[S_{n}^{\alpha},S_{n}^{\beta}\right]\rangle
\end{array}\right)
\end{eqnarray}
We now define everything in terms of a dimensionless \char`\"{}eigenvalue
normalized\char`\"{} effective field $h_{\alpha}=H_{\alpha}/\sqrt{\sum_{\alpha}H_{\alpha}^{2}}$,
and a renormalized frequency $\tilde{\omega}=\omega/\sqrt{\sum_{\alpha}H_{\alpha}^{2}}$.
The explicit calculation of the matrix equation results in: 
\begin{equation}
2\left(\begin{array}{ccc}
0 & ih_{z} & -ih_{y}\\
-ih_{z} & 0 & ih_{x}\\
ih_{y} & -ih_{x} & 0
\end{array}\right)\left(\begin{array}{c}
\langle S_{n}^{\beta}S_{n}^{x}\rangle\\
\langle S_{n}^{\beta}S_{n}^{y}\rangle\\
\langle S_{n}^{\beta}S_{n}^{z}\rangle
\end{array}\right)=\left(\begin{array}{ccc}
\coth\left(\frac{\beta\tilde{\omega}}{2}\right) & -ih_{z} & ih_{y}\\
ih_{z} & \coth\left(\frac{\beta\tilde{\omega}}{2}\right) & -ih_{x}\\
-ih_{y} & ih_{x} & \coth\left(\frac{\beta\tilde{\omega}}{2}\right)
\end{array}\right)\left(\begin{array}{c}
i\epsilon_{x\beta\delta}\langle S_{n}^{\delta}\rangle\\
i\epsilon_{y\beta\delta}\langle S_{n}^{\delta}\rangle\\
i\epsilon_{z\beta\delta}\langle S_{n}^{\delta}\rangle
\end{array}\right)
\end{equation}
In order to solve these coupled equations one must realize that they
are not independent, as the determinant of $\boldsymbol{I}-2\boldsymbol{\Gamma}$
vanishes. The first row multiplied by $h_{x}$, plus the second row
times $h_{y}$, plus the third row times $h_{z}$ gives zero. The
same condition on the right reads: 
\begin{equation}
\epsilon_{x\beta\delta}\langle S_{n}^{\delta}\rangle h_{x}+\epsilon_{y\beta\delta}\langle S_{n}^{\delta}\rangle h_{y}+\epsilon_{z\beta\delta}\langle S_{n}^{\delta}\rangle h_{z}=0
\end{equation}
If we now choose $\beta=x,y,z$ we find, respectively: 
\begin{align*}
\langle S_{n}^{y}\rangle h_{z} & =\langle S_{n}^{z}\rangle h_{y}\\
\langle S_{n}^{z}\rangle h_{x} & =\langle S_{n}^{x}\rangle h_{z}\\
\langle S_{n}^{x}\rangle h_{y} & =\langle S_{n}^{y}\rangle h_{x}
\end{align*}
which are the so called regularity conditions. They imply that we
only need to know one of the components of the magnetization in order
to calculate the other components.

Actually one can solve a somewhat more general system of equations,
again for arbitrary spin. Consider an arbitrary polynomial function
of spin operators of form: 
\begin{eqnarray}
P\left(\left\{ S_{n}^{\alpha}\right\} \right) & = & \sum_{r,p,q=1}^{2S+1}c_{rpq}\left(S_{n}^{x}\right)^{r}\left(S_{n}^{y}\right)^{p}\left(S_{n}^{z}\right)^{q}
\end{eqnarray}
then we have the equation:
\[
\left(\begin{array}{ccc}
0 & ih_{z} & -ih_{y}\\
-ih_{z} & 0 & ih_{x}\\
ih_{y} & -ih_{x} & 0
\end{array}\right)\left(\begin{array}{c}
\langle P_{n}S_{n}^{x}\rangle\\
\langle P_{n}S_{n}^{y}\rangle\\
\langle P_{n}S_{n}^{z}\rangle
\end{array}\right)=\frac{1}{2}\left(\begin{array}{ccc}
\coth\left(\frac{\beta\tilde{\omega}}{2}\right) & -ih_{z} & ih_{y}\\
ih_{z} & \coth\left(\frac{\beta\tilde{\omega}}{2}\right) & -ih_{x}\\
-ih_{y} & ih_{x} & \coth\left(\frac{\beta\tilde{\omega}}{2}\right)
\end{array}\right)\left(\begin{array}{c}
\langle\left[S_{n}^{x},P_{n}\right]\rangle\\
\langle\left[S_{n}^{y},P_{n}\right]\rangle\\
\langle\left[S_{n}^{z},P_{n}\right]\rangle
\end{array}\right)
\]
If we take this set of equations for $P_{n}\equiv P\left(\left\{ S_{n}^{\alpha}\right\} \right)$,
along with the regularity conditions 
\begin{align}
\left(S_{n}^{x}\right)^{2}+\left(S_{n}^{y}\right)^{2}+\left(S_{n}^{z}\right)^{2} & =S\left(S+1\right)\\
\left[S_{n}^{\alpha},S_{n}^{\beta}\right] & =i\epsilon_{\alpha\beta\delta}S_{n}^{\delta}
\end{align}
\begin{align}
\langle S_{n}^{z}\rangle h_{x}=\langle S_{n}^{x}\rangle h_{z},\ \langle S_{n}^{x}\rangle h_{y}=\langle S_{n}^{y}\rangle h_{x},\ \langle S_{n}^{y}\rangle h_{z} & =\langle S_{n}^{z}\rangle h_{y}
\end{align}
and the usual spin algebra identities for spin-$S$ degrees of freedom,
we find that the system of equations for the two-point functions $\langle S_{n}^{\beta}S_{n}^{\alpha}\rangle$
and $\langle S_{n}^{x,y}\rangle$ can be solved as a function of $\langle S_{n}^{z}\rangle$,
i.e., we need one extra equation to solve the system. This can be
obtained from the identity: 
\begin{equation}
\prod_{r=-S}^{S}\left(S_{n}^{z}-r\right)=0
\end{equation}
which clearly becomes more and more complicated as the spin $S$ is
increased.

\textbf{Specific Cases involving electronic and nuclear spins:} As
a first check, we can take $S=1/2$. In that case we find that $\left(S_{n}^{z}\right)^{2}=1/4$,
which provides the extra equation needed for the solution. The system
of equations results in: 
\begin{equation}
\langle S_{n}^{\mu}\rangle=\frac{h_{\mu}}{2}\tanh\left(\frac{\beta\tilde{\omega}}{2}\right),\ \langle\left(S_{n}^{\mu}\right)^{2}\rangle=\frac{1}{4},\ \langle S_{n}^{\mu}S_{n}^{\nu}\rangle=i\frac{\epsilon_{\mu\nu\delta}h_{\delta}}{4}\tanh\left(\frac{\beta\tilde{\omega}}{2}\right)
\end{equation}
which is the expected result from the RPA calculation for a spin $1/2$.

Consider now the case of $S=1$, for which the extra equation reads:
\begin{equation}
\left(S_{n}^{z}\right)^{3}=S_{n}^{z}
\end{equation}
Hence, we must obtain statistical averages for $\left(S_{n}^{z}\right)^{3}$
as well by setting $P_{n}=\left(S_{n}^{x}\right)^{p}\left(S_{n}^{y}\right)^{q}\left(S_{n}^{z}\right)^{r}$
(for this case $p,q,r=0,1,2$ is sufficient) and solving for a larger
system of equations. We can then see how a general rule for arbitrary
spins emerges - this was derived by Wang et al. \cite{Fermion-GF}
- and one finds: 
\begin{equation}
\langle S_{n}^{z}\rangle=\frac{\left[\left(2S+1\right)R-Q_{z}\right]\left(Q_{z}+R\right)^{2S+1}+\left[\left(2S+1\right)R+Q_{z}\right]\left(Q_{z}-R\right)^{2S+1}}{2R^{2}\left[\left(Q_{z}+R\right)^{2S+1}-\left(Q_{z}-R\right)^{2S+1}\right]}
\end{equation}
where we have defined 
\begin{equation}
R=1/\left|h_{z}\right|,\ Q_{z}=\frac{\coth\left(\frac{\beta\tilde{\omega}}{2}\right)}{h_{z}}
\end{equation}
Now it is easy to see how to deal with a set of coupled nuclear and
electronic spins. We first consider the nuclear spin averages. As
an example, take the case where $I_{n}=7/2$ (the case of the $Ho$
nuclear spins in the $LiHoF$ system). Then for the nuclear spin averages
we get 
\begin{align}
\langle I_{n}^{z}\rangle & =\frac{1}{2Q_{z}}+\frac{2Q_{z}}{Q_{z}^{2}+R^{2}}+\frac{8Q_{z}\left(Q_{z}^{2}+R^{2}\right)}{Q_{z}^{4}+6Q_{z}^{2}R^{2}+R^{4}}\\
\langle I_{n}^{y}\rangle & =\frac{h_{y}}{h_{z}}\langle I_{n}^{z}\rangle,\ \langle I_{n}^{x}\rangle=\frac{h_{x}}{h_{z}}\langle I_{n}^{z}\rangle
\end{align}
As these equations depend on $\tilde{\omega}=\tilde{\omega}\left(\langle I_{n}^{x}\rangle,\langle I_{n}^{y}\rangle,\langle I_{n}^{z}\rangle\right)$,
we have to solve them numerically.

If we now consider the full $LiHo$ system, ie., with a coupling between
an Ising electronic system ($S=1/2$) and a spin bath ($I=7/2$),
the self-consistency equations are coupled for all values of the magnetization.
We then find: 
\begin{align}
M_{\mu} & =\frac{h_{\mu}^{S}}{2}\tanh\left(\frac{\beta\tilde{\omega}}{2}\right)\\
m_{\mu} & =\frac{h_{z}^{B}}{2}\left[\tanh\left(\frac{\beta\tilde{\Omega}}{2}\right)+2\tanh\left(\beta\tilde{\Omega}\right)+4\tanh\left(2\beta\tilde{\Omega}\right)\right]\nonumber 
\end{align}
for the electronic system and nuclear bath magnetization respectively,
where $h_{\mu}^{S\left(B\right)}=H_{\mu}^{S\left(B\right)}/\tilde{\omega}\left(\tilde{\Omega}\right)$
represent the system(environment) normalized fields respectively,
and $\tilde{\omega}\left(\tilde{\Omega}\right)$ represent the normalized
eigenenergies of the system(environment) respectively. Note that now
$h_{\mu}^{S}=h_{\mu}^{S}\left(M_{\mu},m_{\mu}\right)$, $h_{\mu}^{B}=h_{\mu}^{B}\left(M_{\mu}\right)$,
$\tilde{\Omega}=\tilde{\Omega}\left(M_{\mu}\right)$ and $\tilde{\omega}=\tilde{\omega}\left(M_{\mu},m_{\mu}\right)$.
We can easily obtain the $T=0$ limit from these expressions; we get:
\begin{align}
M_{\mu} & =\frac{1}{2}\frac{B_{\mu}+M_{\mu}V_{0}^{\mu}-A_{0}^{\mu}m_{\mu}}{\sqrt{\left(B_{x}-A_{0}^{x}m_{x}\right)^{2}+A_{0}^{y}m_{y}^{2}+\left(V_{0}^{z}M_{z}-A_{0}^{z}m_{z}\right)^{2}}}\\
m_{\mu} & =-\frac{7}{2}\frac{A_{\mu}M_{\mu}}{\sqrt{\left(A_{0}^{x}M_{x}\right)^{2}+\left(A_{0}^{y}M_{y}\right)^{2}+\left(A_{0}^{z}M_{z}\right)^{2}}}
\end{align}
We can directly substitute $m_{\mu}$ into the system's magnetization.
Since we are interested in the behavior at large $B_{x}$, in order
to see if the QPT can be blocked, we note that in the asymptotic limit
$B_{x}\gg A_{\mu},V_{0}^{z}$, one finds: 
\begin{equation}
M_{x}\lesssim\frac{1}{2},\ M_{z}\simeq\frac{m_{z}V_{0}^{z}}{2B_{x}}+\frac{7\left(A_{0}^{z}\right)^{2}m_{z}}{4B_{x}\sqrt{\left(A_{0}^{x}m_{x}\right)^{2}+\left(A_{0}^{z}m_{z}\right)^{2}}}
\end{equation}
Clearly in the second equation we could have $M_{z}=0$ as a solution
of the system, meaning that a QPT would exist. However, if we assume
the highly anisotropic case $A_{z}\neq0$, $A_{x}=0$, we find: 
\begin{equation}
M_{z}=\frac{1}{2}\frac{M_{z}V_{0}^{z}+\frac{7}{2}A_{0}^{z}}{\sqrt{B_{x}^{2}+\left(V_{0}^{z}M_{z}+\frac{7}{2}A_{0}^{z}\right)^{2}}}\simeq\frac{M_{z}V_{0}^{z}}{2B_{x}}+\frac{7A_{0}^{z}}{4B_{x}}
\end{equation}
which proves that $M_{z}$ will always have a remnant magnetization
blocking the QPT at $T=0$ for all $B_{x}$ (the second equation can
never be fulfilled when $M_{z}=0$). Hence, the longitudinal hyperfine
coupling blocks the phase transition as one would expect.

\section{Dynamics across the quantum critical point}

Here we include simulations of the magnetization dynamics for the
Quantum Ising model coupled to a spin bath when it crosses a QCP.
The simulation is performed for a slightly simpler version of the
Ising model than the one considered in the main text, but the differences
should not be important for the final conclusions. We consider a finite
Ising system coupled to a spin bath made of $\frac{1}{2}$-spins (we
assumed $I=1/2$ instead of $I=7/2$ for simplicity, but the renormalization
of the hyperfine coupling should make no difference between the two).
We then integrate numerically over time the Heisenberg equation of
motion under the same decoupling scheme used for the calculation of
the Green's functions. Fig.\ref{fig:Annealing1}(left) shows the case
without AC field and hyperfine coupling, as a test for the protocol.
As the transverse DC field increases, the longitudinal magnetization
decreases, vanishing when $B=B_{c}$. The small oscillations around
the mean value $M_{z}=0$ correspond to non-adiabatic effects during
the protocol. Fig.\ref{fig:Annealing1}(right) corresponds to the
case with hyperfine coupling, where the QPT is suppressed due to the
transverse blocking mechanism and a remnant magnetization is always
present. Finally, the case with AC field and hyperfine coupling is
not explicitly shown due to the fast oscillations of the magnetization,
but we find that the average value of the longitudinal magnetization
is $\int_{0}^{T}M_{z}\left(t\right)dt\simeq0.0003$ for $\mathcal{J}_{0}\left(\alpha\right)=0$
and frequency $\omega=10V_{i,j}$. This indicates that the renormalization
of the longitudinal hyperfine coupling persists when all corrections
to the Magnus expansion are included, and it is even possible to adiabatically
cross the QCP.

\begin{figure}
\includegraphics[scale=0.8]{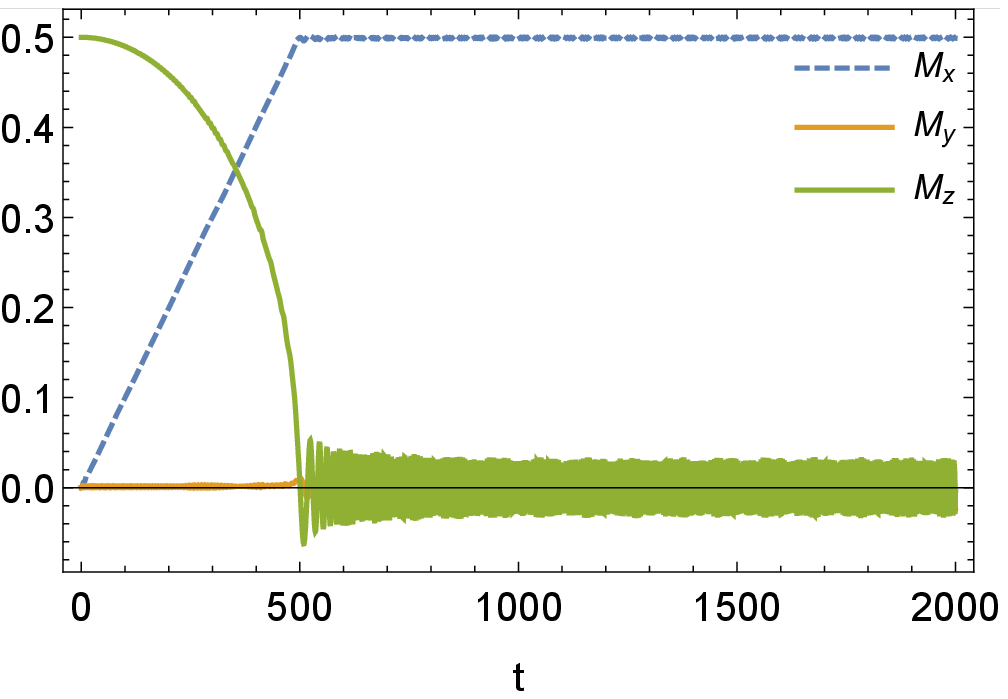}\includegraphics[scale=0.8]{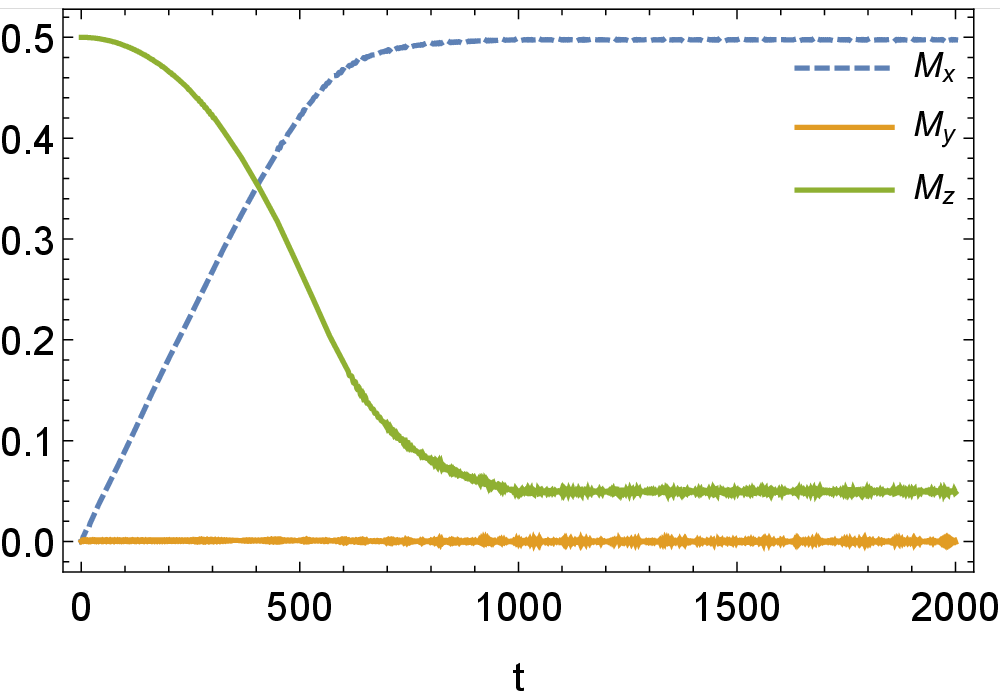}

\caption{\label{fig:Annealing1}(Left) Magnetization as a function of time
for the Quantum Ising model in absence of AC field and hyperfine coupling.
During the protocol, the transverse magnetic field increases linearly
to its final value $B_{x}=2B_{c}$ at $t=1000$. At $t=500$ the system
crosses the QCP and non-adiabatic effects produce the oscillations
that persist at large times (which average to $M_{z}=0$ indicating
the presence of the PM phase). (Right) Simulation in the presence
of hyperfine coupling, where the magnetization remains finite due
to the transverse blocking mechanism, indicating the absence of QPT
to the PM phase. Time is measured in inverse units of $V_{i,j}$.}

\end{figure}

\end{widetext}

\end{document}